\documentstyle[12pt]{article}

\renewcommand{\in}{\raise -3pt\hbox{\scriptsize in}}
\newcommand{\out}{\raise -3pt\hbox{\scriptsize out}}

\newcommand{\beq}{\begin{equation}}
\newcommand{\eeq}{\end{equation}}
\newcommand{\barr}{\begin{eqnarray}}
\newcommand{\earr}{\end{eqnarray}}
\newcommand{\ba}{\begin{array}}
\newcommand{\ea}{\end{array}}

\begin{document}

\begin{titlepage}
\begin{center}
{\Large \bf Positivity constraints for Off-Forward Parton Distributions}\\[1.5cm]
{ \bf B. Pire$^{a}$, J. Soffer $^{b}$ and O.Teryaev$^{a,c}$}\\[0.3cm]

$^a$CPhT\footnote {Unit\'e Mixte de Recherche C7644 du Centre
National de la Recherche Scientifique}, Ecole Polytechnique, F-91128
Palaiseau, France\\
\vskip 0.5cm
$^b$CPT-CNRS\footnote {Unit\'e Propre de Recherche 7061 du
Centre National de la Recherche Scientifique}, F-13288 Marseille Cedex
9, France
\\
\vskip 0.5cm
$^c$Bogoliubov Laboratory of Theoretical Physics,\\
Joint Institute for Nuclear Research\\
141980 Dubna, Moscow region, Russia\footnote{Permanent address}
\end{center}


\begin{abstract}
  Off-Forward Parton Distributions (OFPD's) are new hadronic objects which
may be measured in various exclusive reactions. We derive
non-trivial positivity constraints for them that should allow to get
extra restrictions for model inputs.
\end{abstract}
\noindent Unit\'e Propre de Recherche 7061

\noindent CPT-98/P.3633

\noindent Web address: www.cpt.univ-mrs.fr
\end{titlepage}

1  - The concept of Off-Forward (non-forward, non-diagonal)
Parton Distributions (OFPD),
related to the matrix elements of non-local string operators \cite{OFPD},
has attracted much attention since it has been recognized\cite{Ji,Rad}
that these new objects describing the deep hadronic structure
could be measured in deep exclusive reactions such as forward virtual
Compton scattering and diffractive electroproduction of mesons.
They factorize\cite{Rad,Coll} from a hard subprocess amplitude,
provided the virtuality
$Q^2$ of the photon is large enough for the differential cross section
to enter a scaling regime, where the handbag type diagrams
dominate\cite{DGPR}. As for any long distance dominated object, not
much is known about these distributions, except some limiting values
obtained from already measured standard forward parton distributions.
Various model estimates have been recently proposed\cite{Ji,PPP,Sac,MR,Rad2}, but in the absence of any trustable non-perturbative QCD calculations, we
want here to advocate the usefulness of bounds coming from positivity
requirements for constructing models,
which should allow to get seriously guided rate estimates for
several proposed experiments at CEBAF, CERN and DESY.

For every parton species, there are six off-forward parton distributions.
 They all depend on three kinematical variables, which can be
chosen as $x$, the light-cone fraction of the parton emitted by the proton
target, $x'$, the fraction of the parton absorbed by the scattered
proton, and
$t$, the momentum transfer between the initial and final proton.
Both momentum fractions are measured with respect to the initial proton
momentum $p$ \cite{Rad}. It is meaningful
for positivity studies as well as for symmetry properties \cite{MPW} to
reexpress the light-cone fraction of the parton absorbed by the final
proton as a fraction with respect to that proton's momentum, {\it i.e.}
$x_1 = x$ and $x_2 = x'/(1-x+x')$.
This may be compared with the symmetric choice \cite{Ji},
which, strictly speaking, is understood when OFPD are considered.
We restrict ourselves to the case
when the momentum fraction of the second (absorbed) parton is
positive,
since there is then a clear
relation between the OFPD and ordinary distribution functions
measured in deep inelastic scattering \cite{Rad2}.
Kinematics fixes $t$ and the difference
$x - x'$ to some fixed value like $x_{bj}$ in the deeply virtual Compton
scattering (DVCS) process, while the
scattering amplitude has an imaginary part with $x'=0$ and a real part, which
 is a principal part integral over $x'$.
Here we pay special attention to spin-averaged quark $q_\zeta(x)$ \cite{Rad}
and gluon $g(x,x',t)$ \cite{MR} distributions.

The
$t-$dependence of the OFPD's is governed by the proton form factors
through  relations such as :
$\int_{-1}^{+1} dx_1 ~g(x_1,x_2,t) = F(t)$.
It is reasonable to assume that this $t-$dependence factors out. Remember
however that kinematics fixes $t_{min} \neq 0$. The OFPD's acquire a
$Q^2$-dependence governed by evolution
equations\cite{Ji,Rad,MPW,BGDS,MPS}, and we show at the end of the
paper that the QCD $Q^2$- evolution preserves the validity of the positivity
bound.

\vskip 12pt
\noindent
 2 - Since our present knowledge on OFPD's is rather limited, any
rigorous bounds for them are of great interest.
The aim of the present paper is to develop such bounds, coming from
positivity of the density matrix. Because the OFPD's do not have
a probabilistic interpretation, one may wonder if this is possible at
all. However, non-diagonal elements of a density matrix are
constrained by positivity as well as its diagonal elements,
as shown by the Soffer bound on
the chiral-odd quark distribution $h_1^q(x)$\cite{JJRP}
(this
distribution is forward in momentum, but it is non-diagonal in helicity),
which reads
\cite{Soffer}:
\begin{equation}\label{in}
|h^q_1(x)| \le q_+(x) \equiv  {1 \over 2}[q(x)+\Delta q(x)],
\eeq
where $q$ and $\Delta q$ are the usual spin-averaged and
spin-dependent quark distributions.
Although this interesting result was originally proven at the level of
the parton model, it was shown recently that it is preserved by the
QCD $Q^2$ evolution, up to next-to-leading order\cite{WV,BST}.

The OFPD's are elements of parton density matrices which are
non-diagonal in momentum, and they may be treated in a similar way,
provided the momentum fraction of the absorbed parton is
positive\cite{DG}.  Moreover, in the recent paper \cite{MR} the
inequality
\begin{equation}\label{MR}
2 x^{'} g (x,x^{'}) \le x g(x)+x^{'} g (x^{'})
\eeq
was obtained by a  rather similar method to that of \cite{Soffer}.

Let us now derive another, stronger (especially at
low $x$) inequality, and outline the method allowing to derive
similar inequalities for the various spin components of the OFPD's.
We present here the derivation with some details, in order to
stress the dependence of the actual definition of the non-forward
distribution\footnote{We are indebted to A.V. Radyushkin for pointing
out the correct definitions and for helpful comments,
which allowed to restore the related factors,
which were missing in the original version of this paper.}.

Let us start the discussion with the simpler case of
non-polarized quark distribution, which by introducing
the light-cone decomposition of the quark fields \cite{Jaf}
is analogous to the
scalar one. The quark forward distribution is just
\barr\label{q(x)}
q(x)=\int{d \lambda \over{4\pi}}e^{i\lambda x}<p,S|\bar\psi(0)
/\!\!\! n
 \psi(\lambda n)|p,S>=\nonumber \\
{1 \over{\sqrt{2}p^+}}
\int{d \lambda \over{2\pi}} e^{i\lambda x}<p,S|\phi^+(0)\phi(\lambda n)|p,S>~,
\earr
where $\phi$ is the good component of the quark field and the light-cone
vectors are normalized such as $pn=p^+ n^-=1$. By inserting
a complete set of intermediate states $|X>$
and making use of the generalized optical theorem
and the fact that the matrix elements may be replaced by their
imaginary parts \cite{DG}, the forward
distribution can be written as
\begin{equation}
q (x)=\sum_{X} {1 \over{\sqrt{2}p^+}} |<p,S|\phi(0)|X>|^2
\delta(x-(p-p_X)n)~.
\eeq
The quark non-forward distribution reads,
\begin{equation}\label{qzeta(x)}
q_\zeta (x)={1\over \sqrt{1-\zeta}}
\int {d \lambda \over {4\pi}} e^{i\lambda x}<p,S|\bar\psi(0) /\!\!\! n
\psi(\lambda n)|p',S>~,
\eeq
where the factor $\sqrt{1-\zeta}$ comes from the bilinear $\bar u(p')
u(p)$, in the definition. By an analogous procedure as above, it becomes
\barr
q_\zeta(x)=2 Re \sum_{X} {1 \over {\sqrt{2(1-\zeta)}p^+}}<p,S|\phi(0)|X> \nonumber \\
<p',S|\phi(0)|X>^*\delta(x-(p-p_X)n)~,
\earr
where we used the hermiticity of the matrix element.
We are now ready to write down the Cauchy-Schwarz inequality as:
\barr\label{CS}
\sum_{X}|<p,S|\phi(0)|X> \pm a<p',S|\phi(0)|X>|^2 \delta(x-(p-p_X)n) \geq 0~,
\earr
where $a$ is a positive number, which we put equal to 1 for the time being.
While the non-diagonal term of (\ref{CS}) is producing just the
non-forward distribution, and the first diagonal term - the
distribution $q(x)$, the second diagonal term
should be studied in more details:
\barr
\sum_{X} {1 \over{\sqrt{2}p^+}} |<p',S|\phi(0)|X>|^2
\delta(x-(p-p_X)n)= \nonumber \\
\sum_{X} {1 \over{\sqrt{2}p^+}} |<p',S|\phi(0)|X>|^2
\delta(x'-(p'-p_X)n)= \nonumber \\
\sum_{X} {1 \over{\sqrt{2}p^{'+}}} |<p',S|\phi(0)|X>|^2
\delta[x'(n'^{-}/n^{-}) - (p'-p_X)n'] = q(x_2)~.
\earr
Here the necessary rescaling of the light-cone coordinate, which is
required to get the definition (\ref{q(x)}) is making the argument
equal just $x_2$, while the overall factor $1-\zeta$ coming from the
rescaling of the delta-function argument, is precisely cancelled with
the rescaling of the factor $1/p^+$ (which is the natural consequence
of the correct transformation properties of $q(x)$),
so that the overall rescaling
of the diagonal term \cite{Rad2} is actually manifested for
the scalar case only.

As a result, we have the following inequality
\begin{equation}\label{q(ar)}
|q_\zeta (x)| \le
{1 \over{\sqrt{1-\zeta}}}
[q (x_1)+ q (x_2)]
\eeq
for the spinor case, and
\begin{equation}
|q_\zeta (x)| \le
q (x_1)+ {1 \over{1-\zeta}} q (x_2)
\eeq
for the scalar case. By restoring the dependence on the parameter
$a$, one is led to
\begin{equation}
|q_\zeta (x)| \le
{1 \over{\sqrt{1-\zeta}}}
[a q (x_1)+ {1 \over a} q (x_2)]
\eeq
for spinor quarks and to
\begin{equation}
|q_\zeta (x)| \le
a q (x_1)+ {1 \over{a(1-\zeta})} q (x_2)
\eeq
for scalar quarks. By minimizing the r.h.s. with respect to the
variation of $a$, we finally get \cite{Rad2}
\begin{equation}\label{q(geo)}
|q_\zeta (x)| \le \sqrt
{{ q (x_1) q (x_2)} \over{1-\zeta}}~,
\eeq
for both scalar and spinor quarks.
A similar bound can be obtained for Ji's off-forward quark distribution,that is
\begin{equation}\label{q(geo)}
|H_q(x, \xi)| \le \sqrt
{{ q (x_1) q (x_2)} \over{1-\xi^2}}~,
\end{equation}
where $x_{1,2}=(x \pm \xi)/(1 + \xi), \xi=\zeta/(2-\zeta)$.

The derivation for the gluons is analogous. The forward and nonforward
\cite{Rad,MR} distributions may be expressed as
\barr\label{g)}
xg(x)={1 \over2}\sum_{X,i}|<p,S| G^{+i}(0)|X>|^2\delta(x-(p-p_X)n)~,\nonumber \\ x_2 g (x_2 )={1 \over {2(1-\zeta)}}\sum_{X,i}|<p',S|G^{+i}(0)|X>|^2\delta(x-(p-p_X)n)~,\nonumber \\
x'g(x_1,x_2)={1 \over{\sqrt{1-\zeta}}}Re \sum_{X}<p,S|G^{+i}(0)|X> \nonumber \\
<p',S|G^{+i}(0)|X>^*\delta(x-(p-p_X)n)~,
\earr
where the summation over $i$ stands to select the transverse components of the gluon field of strength $G$. The Cauchy-Schwarz inequality leads to
\begin{equation}\label{g(ar)}
|x^{'} \sqrt{1-\zeta} g(x_1,x_2)| \le \frac{1}{2} [x_1 g(x_1) +
(1-\zeta)
x_2 g(x_2)]~,
\eeq
and, after minimization
with respect to the variation of $a$, one is led to
\begin{equation}\label{g(geo)}
|x^{'} g(x_1,x_2)| \le \sqrt {x_1 x_2 g(x_1) g(x_2)}~.
\eeq
Equality (\ref{g(ar)}) for $x_2 \sim x^{'} \ll x$
is numerically close to (\ref{MR}).
However, the symmetry properties for the variables $x_1, x_2$ are
simpler
\begin{eqnarray}
x'(x_1,x_2) g(x_1,x_2) = x'(x_2,x_1) g(x_2,x_1)~.
\label{m}\end{eqnarray}
It was stressed \cite{Rad2} that in the case of the double
distribution this symmetry is manifested,
provided the overall factor $1-\zeta/2$ \cite{MPW} is extracted.
However, the symmetry of nonforward distribution is more complicated
in that case and one can see that
\begin{eqnarray}
F_\zeta(X)={1 \over{1-\zeta}} F_{-{{\zeta}\over {1-\zeta}}}
({{X-\zeta} \over{1-\zeta}})~.
\label{sf}\end{eqnarray}

Some comments are in order. First, all the considered inequalities
are also valid, when the $t$ dependence of the OFPD's is present in the l.h.s.,
while this dependence is absent in the r.h.s. due
to Lorentz invariance (c.f. \cite{CM}).
Second, the $x$ dependence of the two terms in (\ref{q(ar)})
is not governed by Lorentz invariance, as the light-cone direction
is crucial.

\vskip 12pt
\noindent
3 - Let us now take into account the spin degrees of freedom.
To do so, we consider the quantities
$<p,S| G^{+i} (0) \pm \tilde  G^{+i} (0)|X>$, corresponding
to a definite gluon helicity, while the hadron helicities
are fixed to be positive,
leading to the absence  of the contributions which are
non-diagonal in helicity indices. By applying the same method
as above, one easily gets
\begin{equation}\label{ars}
2 x^{'} \sqrt{1-\zeta} |g^{\pm}(x_1,x_2)| \le x_1 g^{\pm}(x_1)+x_2 (1-\zeta)
g^{\pm}(x_2)~.
\eeq
By adding these two inequalities, one checks that (\ref{g(ar)}) is
still valid, so that unpolarized distributions are decoupling from the
polarized ones. This is no more valid in the case of the optimized
inequalities
\begin{equation}\label{geos}
x^{'} |g^{\pm}(x_1,x_2)| \le \sqrt {x_1 x_2 g^\pm(x_1) g^\pm(x_2)}~,
\eeq
leading to the bound
\begin{equation}\label{geos1}
x^{'} |g (x_1,x_2)| \le \sqrt {x_1 x_2 g(x_1)g(x_2)}\cdot
\lambda [P(x_1),P(x_2)]~,
\eeq
with $2\lambda [P(x_1),P(x_2)]=
\sqrt{(1+P(x_1))
(1+P(x_2))}+\sqrt{(1-P(x_1))(1-P(x_2))}$
, where one introduces the gluon polarization, defined as
$P(x)$=$ \Delta G (x)/G(x)$ and such as $|P(x)| \le 1$.
This inequality, in principle, offers a possibility of extracting
information on the gluon spin-dependent distribution $ \Delta G$
from the unpolarized diffractive processes.
Conversely, if one knows $ \Delta G$
one gets an inequality which is stronger than (\ref{g(geo)})
since one has always the inequality  $\lambda [P(x_1),P(x_2)] \le 1$.

The inequality (\ref{g(geo)}) in turn
provides a stronger bound on $g(x_1,x_2)$,
in comparison with (\ref{g(ar)}), and this is related to the difference
between $g(x_1)$ and $g(x_2)$.
It is especially pronounced when one of the $x$
is small, a situation occuring in diffractive electroproduction.
At the same time, a bound for the behaviour of the OFPD's in
the quasielastic region $x_1 \to 1, x_2=const.$ is implied by the stronger
inequality (\ref{g(geo)}), while it cannot be derived from the weaker one.
Namely, the OFPD's should decrease like $(1-x_1)^{\beta/2}$, where the power
$\beta$ characterizes the decrease of the forward distribution and is
related to the form factor behaviour by the quark counting rules.
In particular, the ratio $R$ defined in \cite{MR}, as $x'g(x,x')/xg(x)$
is bounded as
\begin{equation}\label{R}
R \le
\sqrt {x'/(1-x+x') g(x'/(1-x+x')) \over x g(x)} \approx
\sqrt {{{x^{'} g(x^{'})} \over {x g(x)}}}~,
\eeq
where we neglected the difference between $x^{'}$ and $x_2$.

For a better estimate one may use the parametrization $g(x)= \\  Nx^{-\alpha} (1-x)^\beta$,
so that the growth of R for small $x,x^{'}$ is bounded as
\begin{equation}\label{RP}
R \le (x / x^{'})^{{\alpha-1}\over2}~,
\eeq
while the power is twice larger for the weaker bound.

\vskip 12pt
\noindent
4- Up to now, we have considered non-diagonality either in helicity
\cite{Soffer} or in momentum. It is also possible to consider both
effects together, by a simple generalization of the outlined
method. Let us consider in the quark sector
the distributions $q_+(x)$ and $h(x,x^{'})$,
the latter being the non-forward generalization of transversity
\cite{JH}. By optimization with respect to $a$,
one gets the obvious non-trivial
 bound
\begin{equation}\label{hnf}
 |h (x_1,x_2)| \le \sqrt{q_+(x_1)q_+(x_2)}~.
\eeq
One may derive other inequalities, considering various combinations
of the quantities $a$, and varying the helicity indices in
their definitions.

\vskip 12pt
\noindent
5- To check the validity of the positivity bounds in the case of the
leading order $Q^2$ evolution, one may use the kinetic interpretation
of the latter \cite{BLT,BST}, similarly to the proof for the
Soffer inequality.
As a result, one finds that the positivity constraint
(\ref{g(geo)}) is preserved provided the
following inequality is satisfied
\begin{equation}\label{ker}
 \sqrt{{{z'(1-z)} \over {z(1-z')}}}
P (z,z') \le \sqrt{P(z)P(z')}~,
\eeq
where $P(z)$ and $P(z,z')$ are the non-singular parts of the
diagonal and off-diagonal \cite{MR} splitting kernels.
Note that the factor $\sqrt{{{z'(1-z)} \over {z(1-z')}}}$ makes
the l.h.s. symmetric with respect to the  interchange $z
\leftrightarrow z'$. Since $P(z,z')$ is itself
symmetric with respect to the transformation $z \to 1-z', z' \to
1-z$, the l.h.s. is also symmetric with respect to the
simultaneous interchange $z
\leftrightarrow 1-z, z' \leftrightarrow 1-z'$, while the symmetry
with respect to these interchanges made separately, also respected by
r.h.s., is violated by the factor $1/2$ of the two last terms in
$P(z,z')$. One easily
checks that (\ref{ker}) is actually satisfied by the kernel \cite{MR}.
In the next-to-leading order case, the positivity is
dependent on the factorization scheme and may be used \cite{BLT}
, as an extra constraint, for making the suitable choice.

\vskip 12pt
\noindent
In conclusion, let us stress that the positivity constraints derived
here will help model builders to improve their rate
estimates for proposed electroproduction experiments.

\vskip 12pt
\noindent
{\bf Acknowledgements}
We acknowledge useful discussions and correspondence
with M. Diehl, I.V. Musatov, A.V. Radyushkin and J.P. Ralston.
This investigation was
partially supported by grants 96-02-17361 from  Russian Foundation
for Fundamental Research and 93-1180ext from INTAS.

\end{document}